\documentclass[12pt]{article}
\newfont{\bbd}{msbm10 scaled\magstep1}
\def\C{\hbox{\bbd C}}
\def\beq{\begin{equation}} \def\eeq{\end{equation}}
\def\be{\begin{displaymath}} \def\ee{\end{displaymath}}
\def\bea{\begin{eqnarray}} \def\eea{\end{eqnarray}}
\def\beas{\begin{eqnarray*}} \def\eeas{\end{eqnarray*}}
  \def\phi{\varphi}

 \def\a{\alpha}

\def\Ref#1{(\ref{#1})}

\newcommand{\tfrac}[2]{{\textstyle\frac{#1}{#2}}}

\begin{document}
\renewcommand{\thefootnote}{\fnsymbol{footnote}}
\begin{center} {\Large \bf Simultaneous separation for the 
Kowalevski and Goryachev-Chaplygin
gyrostats}
\end{center}
\renewcommand{\thefootnote}{\arabic{footnote}}
\vskip1cm
\renewcommand{\thefootnote}{\fnsymbol{footnote}}
\begin{center} {\bf Vadim B.~Kuznetsov\footnote[2]{EPSRC 
Advanced Research Fellow}}
\end{center}
\renewcommand{\thefootnote}{\arabic{footnote}}
\vskip 0.5cm
\begin{center} Dept of Applied Maths, University of Leeds, LEEDS LS2 9JT, 
United Kingdom\\
E-mail: vadim@maths.leeds.ac.uk
\end{center}
\vskip2cm
\begin{center}
{\bf Abstract} \end{center}
\vskip0.2cm

\noindent
In the special case of zero square integral
the Kowalevski gyrostat and Goryachev-Chaplygin gyrostat
share a simple separation of variables originated from the 
$4\times 4$ Lax matrix.

\vskip 5cm

\noindent
\vskip 5cm

\pagebreak
%
%%%%%%%%%%%%%%%%%%%%%%%%%%%%%%%%%%%%%%%%%
%
\section*{Introduction}
The theory of {\it separation of variables} (SoV) is the theory of special canonical 
transformations and the theory of {\it quantum separation of variables} is the theory 
of corresponding integral transforms. The former was already understood in this sense
starting from the development of the Hamilton-Jacobi approach for solving 
Liouville integrable systems. The real power of the latter, which by now is undoubted,
requires further investigation and demonstration.

The {\it speciality} of a separating transformation stems from its definition as a transform
resulting in new variables {\it being separated}  or equations {\it 
being decoupled} from one another. Below we give a (working) definition 
of SoV in the context of finite-dimensional integrable Hamiltonian dynamics.

By {\it separation of variables} for an $n$-degrees-of-freedom integrable system
having $n$ independent Poisson commuting integrals $\{H_j({\bf q},{\bf p})\}_{j=1}^n$
we mean a canonical transformation 
from the (old) Darboux variables $q_j,p_j$, $j=1,\ldots,n$, to new Darboux
variables $u_j,v_j$, $j=1,\ldots,n$, which satisfy the following 
{\it separation equations}:
\begin{equation}
\sum_{j=1}^n a_{ij}(u_i,v_i)\,H_j({\bf q},{\bf p})=b_i(u_i,v_i),\qquad i=1,\ldots,n.
\label{se-1}
\end{equation}
In other words, being expressed in terms of the new variables,
the integrals of motion $H_j$ acquire 
the following `separated form':
\begin{equation}\label{sf}
{\bf H}={\bf A}^{-1} {\bf B},\qquad ({\bf A)}_{ij}=a_{ij},\qquad ({\bf B})_i=b_i,\qquad ({\bf H})_i=H_i.
\end{equation}
The conditions that the functions $a_{ij}$ and $b_i$ in (\ref{se-1}) depend
on the new (separation) variables with the index $i$  {\it only}, is crucial. It indeed
means that the $n$ equations in (\ref{se-1}) are really separated from one another.
Therefore, they are $n$ equations, each of one degree of freedom, sharing only the 
common values of the Hamiltonians $H_j$ \footnote{as one can see, this definition
is similar to the one defining St\"ackel systems}.

The above definition includes, as even more special canonical transform,
the (classical) coordinate separation of variables
when the new coordinates, say ${\bf u}$, are the functions of the
old coordinates (${\bf q}$) only, and they do not depend on the momenta 
(${\bf p}$). See \cite{Ka,Ku92a,Ku92b,EEKT} for
some history of this sub-class of transformations and for many examples
of such situation. 

General separating canonical transforms, however, have
new (separation) variables which are non-trivial functions of all $2n$
initial canonical variables,
\begin{equation}\label{ss-exx}
u_j=u_j({\bf q}, {\bf p}),\qquad
v_j=v_j({\bf q}, {\bf p}),\qquad j=1,\ldots,n,\qquad
\end{equation}
and which satisfy the standard Poisson brackets
\begin{equation}
\{u_j,u_k\}=\{v_j,v_k\}=
\{u_j,v_k\}=0,\qquad j\neq k,\qquad \{v_j,u_j\}=1,\qquad j=1,\ldots,n.
\end{equation}
Examples of such separating canonical transforms are usually
much more sophisticated than those of the coordinate ones. As far as I know,
the first explicit example was given by van Moerbeke in 1976 in \cite{M76}
concerning the separation of variables for the periodic Toda lattice
(see also \cite{FM}). In 1980 Kozlov \cite{Kozlov} rewrote 
the classical results of Goryachev and Chaplygin on integration in
quadratures of the Goryachev-Chaplygin top as a simple canonical transform.
The method of SoV, viewed as a method of special canonical transformations,
was further developed by Komarov in a series of works on tops, including
quantum separation of variables, see \cite{K81,Kom,Kom2}. Many further examples 
have been
produced since 1982, with the theory benefiting mostly from the developments
of the algebraic geometric and $r$-matrix understanding of the 
method of separation of variables. This led to a rather satisfying picture
of the present state-of-art of non-coordinate separation of variables.
See \cite{Skl} and \cite{Hu,Hu2} for more details.

Notice here that in the realm of the algebraic geometry usually associated with many
Liouville integrable systems, the separation equations (\ref{se-1}) appear 
as equations for a set of separating Darboux coordinates on the 
spectral curve of a Lax matrix $L(u)$, namely:
\begin{equation}
\mbox{(\ref{se-1})}\qquad \Leftrightarrow \qquad\det(L(u_i)-v_i)=0,\quad i=1,\ldots,n.
\end{equation}
SoV is not unique, so that
in the situation when there exist several Lax matrices for
the same integrable system, one should expect several 
separations. But even for the same Lax matrix there are many
different separations given by different sections
\footnote{those usually differ by the number of non-moving poles of the 
Baker-Akhiezer function}, cf. \cite{KNS,KS98,KV}.

In the present paper we find a new canonical transform
which simultaneously separates two classical tops: the Kowalevski top \cite{SK} and
the Goryachev-Chaplygin top \cite{Gor,Chap}. The latter top is integrable only for the zero 
value of the square integral, $\ell=0$ (which is one of the Casimirs), so that although
the canonical transform will be defined for arbitrary value of $\ell$, it will
separate the Kowalevski top (and the other one) only for $\ell=0$. SoV for the 
Kowalevski top for $\ell\neq0$, which would generalize the found transform,
remains an unsolved problem. 

Sretenskii \cite{Sre} discovered an integrable extension of the Goryachev-Chaplygin
top adding the gyrostatic term to the Hamiltonian. Komarov \cite{Kom87} and, independently,
Yehia \cite{Yehia} found the gyrostat extension of the Kowalevski top in 1987. 
The quantum Goryachev-Chaplygin gyrostat was treated in \cite{Kom84}.
For quantum Kowalevski gyrostat see \cite{Kom87}.
Here we will always include such gyrostatic terms (cf. the parameter $c$ below).

In Section 1 we give definitions of the integrable systems concerned
and their Lax matrices. In Section 2 we recall the separation method
with main results presented in Section 3. In Section 4 we re-write
the found separating transformation through the generating 
function. Solution of the inverse problem is given in Section 5.
Finally, some concluding remarks can be found in the last Section.

%
%%%%%%%%%%%%%%%%%%%%%%%%%%%%%%%%%%%%%%%%%
%
\section{Algebra e(3), tops and Lax matrices}

The Poisson brackets for the e(3) generators $J_k$, $x_k$, $k=1,2,3$, 
are defined in the standard way:
\begin{equation}\label{pb}
\{J_k,J_l\}=\varepsilon_{klm} J_m, \qquad\{J_k,x_l\}=\varepsilon_{klm} x_m,\qquad
\{x_k,x_l\}=0,
\end{equation}
where $\varepsilon_{klm}$ is the completely anti-symmetric tensor, 
$\varepsilon_{123}=1$.

The Casimirs of the bracket (\ref{pb}) have the form
\begin{equation}\label{c}
C_1=x_1J_1+x_2J_2+x_3J_3=\ell,\qquad C_2=x_1^2+x_2^2+x_3^2=1.
\end{equation}

The Kowalevski gyrostat has the Hamiltonian $H$,
\begin{equation}\label{h}
H=J^2+(J_3+c)^2-2bx_1,
\end{equation}
and the second integral $K$,
\begin{equation}
K=(J_3+c)^2J^2+2b(J_3+c)(J_1x_3-J_3x_1)-b^2x_2^2-2b\ell J_1,
\end{equation}
which are Poisson commuting:
\begin{equation}\label{0}
\{H,K\}=0.
\end{equation}
Here we use the following notation for the square of the vector of 
angular momentum:
\begin{equation}\label{0-1}
J^2:=J_1^2+J_2^2+J_3^2.
\end{equation}
The two integrals of motion, $H$ and $K$, define what is called Kowalevski gyrostat.
It is a Liouville integrable system with two degrees of freedom.

Another integrable system closely related to the Kowalevski gyrostat is the 
Goryachev-Chaplygin gyrostat. It is integrable only when $\ell=0$ and is
defined by two Poisson commuting integrals of motion:
\begin{eqnarray}
\hat H&=&J_1^2+J_2^2+(2J_3+c)^2-4bx_1,\\
\hat K&=&(J_3+c)(J_1^2+J_2^2)+2bx_3J_1.
\end{eqnarray}

The Kowalevski top without the gyrostatic term ($c=0$) was integrated by Kowalevski
in 1889 \cite{SK}. See also \cite{Ku02} where a $2\times 2$ Lax matrix was constructed
for the Kowalevski top which is related to the separation of variables
hidden in Kowalevski's integration. There is a large body of
literature dedicated to the Kowalevski
top, including the study of its geometry. See \cite{K81,Kom87,KK87,HH,RS,
AM88,HM89,BRS,KT89a,KK90,Van,Marshall,LM,K00,Mar,Mar2} 
to name just a few references. 
The Goryachev-Chaplygin top and gyrostat also
received much attention in the literature starting from the discovery of this
integrable top by Goryachev and Chaplygin in 1900. We refer the reader
to the works \cite{Gor, Chap, Kozlov, Kom, Kom2, Kom84, S-old,
Bech, Gavrilov, Ku88,KK89}.

In parallel with the notations $J_1$, $J_2$ and $x_1$, $x_2$ we will be using
their equivalent complex versions $J_+$, $J_-$ and $x_+$, $x_-$:
\begin{equation}\label{0-2}
J_\pm=J_1\pm \mbox{\rm i} J_2,\qquad x_\pm=x_1\pm \mbox{\rm i} x_2,
\end{equation}
with the e(3) Poisson brackets (\ref{pb}) replaced, correspondingly, by
\begin{equation}\label{pb-1}
\{J_3,J_\pm\}=\mp \mbox{\rm i} J_\pm, \qquad\{J_+,J_-\}=-2\mbox{\rm i}J_3,\qquad
\{J_3,x_\pm\}=\{x_3,J_\pm\}=\mp \mbox{\rm i} x_\pm,
\end{equation}
\begin{equation}
\{J_+,x_-\}=\{x_+,J_-\}=-2\mbox{\rm i}x_3,\qquad \{J_3,x_3\}=\{J_+,x_+\}=\{J_-,x_-\}=0,
\end{equation}
\begin{equation}
\{x_k,x_l\}=0,\qquad k,l=\pm,3,
\end{equation}
and the Casimirs (\ref{c}) by
\begin{equation}\label{c-1}
2C_1=x_+J_-+x_-J_++2x_3J_3=2\ell,\qquad C_2=x_+x_-+x_3^2=1.
\end{equation}

A $4\times4$ Lax matrix for the Kowalevski gyrostat was found in \cite{RS}
and used in \cite{BRS} to integrate the problem in terms of Prymian theta-functions.
This integration is different from the one performed by Kowalevski. 
It is not obvious how to construct a separation of variables 
related to such Lax matrix. With the general case
still being an interesting and chal\-lenging problem, we show here how 
to do it in the special case, when $\ell=0$. That is, for this special case
we construct a new separation of variables for the Kowalevski top (and
simultaneously for the Goryachev-Chaplygin top)  with separation
variables belonging to the spectral curve of the $4\times4$ Lax matrix
from \cite{RS}. 

The required Lax matrix is as follows:
\begin{equation}\label{laxx}
L(u)=-\mbox{\rm i}
\pmatrix{\frac{c}{u} & -\frac{bx_-}{u^2}& \frac{J_-}{u}& -\frac{bx_3}{u^2}\cr
                      \frac{bx_+}{u^2}& -\frac{c}{u}&\frac{bx_3}{u^2}&-\frac{J_+}{u}\cr
                      \frac{J_+}{u}& -\frac{bx_3}{u^2}&\frac{2J_3+c}{u}&2+\frac{bx_+}{u^2}\cr
                      \frac{bx_3}{u^2}&-\frac{J_-}{u}&-2-\frac{bx_-}{u^2}&-\frac{2J_3+c}{u}}.
\end{equation}
The spectral curve $\Gamma$: $\det(L(u)-v)=0$, of the Lax matrix (\ref{laxx}) has the form

\begin{equation}\label{cucu-1}
\Gamma:\quad
\left(  v^2+\frac{H}{u^2}-\frac{b^2}{u^4} \right)^2
-4\,\frac{u^4v^2+K}{u^4}-4\,\frac{c^2u^4-b^2\ell^2}{u^6}=0,
\end{equation}
or,
\begin{equation}\label{cucu-2}
v^4-2v^2\left(2-\frac{H}{u^2}+\frac{b^2}{u^4}\right)-\frac{4c^2}{u^2}+\frac{H^2-4K}{u^4}
+\frac{2b^2(2\ell^2-H)}{u^6}+\frac{b^4}{u^8}=0.
\end{equation}

As I noticed in \cite{Ku89} (cf. also \cite{Ku88}) this Lax matrix contains the $3\times 3$ Lax matrix 
${\hat L}(u)$ for the Goryachev-Chaplygin top as its $(1,1)$-minor:
\begin{equation}\label{laxxx}
{\hat L}(u)=-\mbox{\rm i}
\pmatrix{ -\frac{c}{u}&\frac{bx_3}{u^2}&-\frac{J_+}{u}\cr
               -\frac{bx_3}{u^2}&\frac{2J_3+c}{u}&2+\frac{bx_+}{u^2}\cr
               -\frac{J_-}{u}&-2-\frac{bx_-}{u^2}&-\frac{2J_3+c}{u}}.
\end{equation}
The spectral curve $\hat\Gamma$: $\det({\hat L}(u)-v)=0$, 
of the Lax matrix (\ref{laxxx}) has the form
\begin{equation}\label{cucu-3}
v^3-\mbox{\rm i}\,\frac{cv^2}{u}-\left(4-\frac{\hat H}{u^2}+\frac{b^2}{u^4}\right)v+
\frac{4\mbox{\rm i}c}{u}+\mbox{\rm i}\frac{2\hat K-c\hat H}{u^3}+
\mbox{\rm i}b^2\,\frac{c+2x_3\ell}{u^5}=0.
\end{equation}
The Lax matrix (\ref{laxxx}) (and the curve (\ref{cucu-3})), 
for $\ell=0$, was used in \cite{Ku88} to construct explicit
theta-function formulas solving the dynamics of the Goryachev-Chaplygin top.

In the present paper I take one step further in studying a close connection between two
problems and show that there exist separation variables
$u_j,v_j$, $j=1,2$, which are canonical and which, for $\ell=0$, belong to both curves, 
$\Gamma$ and $\hat \Gamma$, simultaneously:
\begin{equation}\label{cucu-4}
\left\{\matrix{\det(L(u_j)-v_j)=0,\cr \det(\hat L(u_j)-v_j)=0,}\qquad j=1,2,
\right.
\end{equation}
\begin{equation}
\{u_j,u_k\}=\{v_j,v_k\}=
\{u_j,v_k\}=0,\qquad j\neq k,\qquad \{v_j,u_j\}=1,\qquad j=1,2.
\end{equation}
Therefore, I construct a new separation of variables which is characterized 
by the property of being a {\it simultaneous separation for both tops}.

\section{The method of SoV}

Separation variables had been generally used to construct closed
expressions for the action variables (in terms of abelian integrals)
or to get a separated representation for the action function.
Therefore, the SoV method had served for a long time
an important but technical role in solving Liouville 
integrable systems of classical mechanics.
A new and much more exciting application of the method came with
the development of quantum integrable systems. Because of the fact 
that quantization of the 
action variables seemed to be a rather formidable task, quantum separation
of variables became an inevitable refuge. In fact, it has been successfully
performed for many families of integrable systems (see, for instance,
survey \cite{Skl}).

Starting from about 1982 the method of separation of variables 
gets connected with the $R$-matrix formalism of the quantum
inverse scattering method, developed during that
time by the Leningrad School. It was noticed by Komarov (see \cite{S-old} 
and \cite{Skl}) that for the $2\times 2$
$L$-operators (Lax matrices) the separation 
variables ought to be the zeros of the off-diagonal element
of the $L$-operator. This observation was fully exploited
by Sklyanin in \cite{S-old,S-Toda} who developed a
beautiful (pure algebraic) setting for the method within the framework
of the $R$-matrix technique. Since then this approach took off
and led to separations for many families of integrable systems.
The method was further generalized to include higher rank
$L$-operators and non-standard normalizations,
see the 1995 review \cite{Skl} and the later developments in
\cite{KKM,KS95,KS96,Ku97,KNS,KS98,KS99a,KS99b}.

An alternative, algebraic geometric approach,
which dates back to Adler and van Moerbeke \cite{AM,AM2} and 
Mumford \cite{M} and includes many researchers,
have been developed starting from about the same time
(see, for instance, \cite{Pr,HM89,AHH2,AHH,Van,Hu,Hu2}). 
It is based on thorough studies of the geometry of lower
genus (algebraic completely) integrable systems. It has also
led to many important new separations for complicated
systems and tops.

Below we recall the most important formulas of the SoV method, adopting
the algebraic description and following mainly the work \cite{KNS}
(see also \cite{Skl,Ku97}).

A Baker-Akhiezer function $f$ is the eigenvector of the Lax matrix for
an integrable system
\beq
L(u)\;f=v\;f,
\label{ba}\eeq
considered as a function on the spectral curve $\Gamma:$ $\det(L(u)-v)=0$.
The inverse scattering method pins down the separation variables as 
poles of this function (see the first example of this general fact in \cite{M76}
and \cite{FM}). 
There is, however, a large freedom of similarity
transformations of the Lax matrix,
\beq
L(u)\mapsto V L(u) V^{-1},
\label{ba1}\eeq
which do not change the spectrum of $L(u)$ but change the divisor
of poles of $f$. This freedom can be characterized, and therefore fixed,
by introducing a {\it normalization} of the Baker-Akhiezer function,
\beq
\vec\a\cdot f\equiv 
\sum_{i=1}^N\a_i\;f_i=1\,,\qquad (\;f\equiv
(f_1,\ldots,f_N)^t\;)\,,
\label{5.6}\eeq
which is given by a {\it normalization} (row-) {\it vector} $\vec \alpha=(\alpha_1,\ldots,\alpha_N)$. 
In other words, one consi\-ders a {\it section of the line-bundle} (cf. \cite{Hu,Hu2,KV}).
A proper (separating)
normalization/section should give a divisor ${\cal D}=\sum_{j=1}^n (u_j,v_j)$
of moving poles of $f$, consisting of $n$ (independent) points on the curve 
$\Gamma$ whose
coordinates are canonical variables. The canonicity of the separation
variables is usually checked by a calculation involving a $r$-matrix, but
in lower genus situations it can be proved in a direct calculation.
For a non-separating normalization/section the divisor ${\cal D}$ will usually
have more points than needed, which will not give canonical variables.

Many families of Lax matrices have a simple separating normalization
when one of the components of the Baker-Akhiezer vector is put 1
(and, hence, one looks at the common poles of the other components),
which corresponds to the following vector $\vec\a$ 
\footnote{or to any other similar vector with the 1 elsewhere}:
\beq
\vec\a_0\equiv (1,0,\ldots,0,0)\,.
\label{eq:a0}\eeq
We will call such normalization the {\it standard normalization}. 
There are examples of non-standard
(dynamical) separating normalizations for the systems with elliptic $r$-matrix 
\cite{Skl,Hu2},
Calogero-Moser systems \cite{KNS} and the $D$-type Toda lattice \cite{Ku97}.
See also \cite{KV} where non-standard dynamical separating normalizations
were used to construct B\"acklund transformations. Generally, Lax matrices with
extra symmetries require non-standard separating normalizations.

Now, assuming that we know a separating normalization $\vec\alpha$ for an integrable
system with the Lax matrix $L(u)$, let us derive the equations for the separation
variables $(u_j,v_j)$, $j=1,\ldots,n$.

{}From the linear problem \Ref{ba} and normalization \Ref{5.6} we derive that
$\vec\a\cdot L^k\;f=v^k\,,\;k=0,\ldots,N-1$, hence,
\beq
f= \left( \begin{array}{c} \vec{\alpha}\cr \vec{\alpha}\cdot L(u) \cr
\vdots \cr \vec{\alpha}\cdot L^{N-1}(u) \end{array}\right)^{-1} \cdot 
\left(\begin{array}{c} 1\cr v\cr  \vdots \cr v^{N-1}\end{array}\right)\,.
\label{NN1}\eeq

Another useful representation for the eigenvector $f$, which can be
verified directly, is as follows:
\beq
f_j=\frac{(L(u)-v)^\wedge_{jk}}{(\vec\a\cdot(L(u)-v)^\wedge)_k}\,,
\qquad \forall k=1,\ldots,N\,,
\label{NN2}\eeq
where the wedge denotes the adjoint matrix. It follows from it that the poles 
of $f$ are the common zeros of the vector equation:
\beq
\vec \a\cdot (L(u)-v)^\wedge=0\,.
\label{5.11}\eeq
Eliminating $v$ from these equations, one can get a single equation
for the $u$-components of the separating variables as zeros of the
following determinant:
\beq
B(u)=\det\pmatrix{\vec\a\cr
\vec\a \cdot L(u)\cr
\vdots\cr
\vec\a \cdot L^{N-1}(u)}=0\,.
\label{5.12}\eeq
Notice that this is exactly the denominator in the representation (\ref{NN1}).

Also, from the equations (\ref{5.11}) we can obtain explicit formulas for the $v$-components
of the separation variables in the form
\beq
v=A(u),
\label{vvv}\eeq
with $A(u)$ being rational functions of the entries of $L(u)$.
Let us derive those formulas. 

Define the matrices $L^{(p)}$, $p=1,\ldots,N$, with the following entries:
\beq
L^{(p)}_{ij}:=\sum_{i_1=1}^N\cdots \sum_{i_{p-1}=1}^N\;\left| 
\begin{array}{cccc}
L_{i,j}&L_{i,i_1}&\cdots & L_{i,i_{p-1}} \\ 
L_{i_1,j}&L_{i_1,i_1}&\cdots & L_{i_1,i_{p-1}} \\ 
\vdots &\vdots & \ddots & \vdots \\ 
L_{i_{p-1},j}& L_{i_{p-1},i_1} &\cdots & L_{i_{p-1},i_{p-1}}
\end{array}\right|\,,\quad p=2,3,\ldots,N\,,   
\label{eq:expA}\eeq
and put $L^{(1)}\equiv L$. These matrices satisfy the 
recursion relation of the form
\beq
L^{(p)}= L\left({\rm tr}\,L^{(p-1)}\right) - (p-1)\, L^{(p-1)} L\,.
\label{eq:recurs}\eeq 
Introduce the matrix ${\cal B}(u)$ by the formula
\beq
{\cal B}(u):=\pmatrix{
          \vec\a\cdot L^{(1)}(u)\;L^{-1}(u)\cr
          \vec\a\cdot L^{(2)}(u)\;L^{-1}(u)\cr
\tfrac12\;\vec\a\cdot L^{(3)}(u)\;L^{-1}(u)\cr
\cdots\cr
\tfrac{1}{(N-1)!}\;\vec\a\cdot L^{(N)}(u)\;L^{-1}(u)}\,.
\eeq
With the help of this matrix we can represent the system
of equations (\ref{5.11}) as a system of linear homogeneous equations
for the vector of powers of $(-v)$:
\beq
\vec\a\cdot(L(u)-v)^\wedge
\equiv((-v)^{N-1},(-v)^{N-2},\ldots,1)\cdot\,{\cal B}(u)=0,
\eeq
from which we derive that 
\beq
(-v)^{j-i}=\frac{({\cal B}^\wedge(u))_{ki}}{({\cal B}^\wedge(u))_{kj}}\,,
\qquad \forall k\,.
\label{super}\eeq
The formulas \Ref{super} give plenty of representations for the rational
functions $A(u)$ in (\ref{vvv}), all of them being compatible on the separation variables since,
because of the equality
\beq
B(u)=(-1)^{[N/2]}\det({\cal B}(u))\,,
\eeq
the matrix ${\cal B}^\wedge(u_j)$ has rank 1. For more details see \cite{KNS}.

\section{A new separation of variables}
Consider the standard normalization vector
\beq
\alpha_0=(1,0,0,0)
\label{f-1}\eeq 
for the $4\times 4$ Lax matrix (\ref{laxx}). Then, for $\ell=0$, the defining equations,
\beq
\left(L(u)-v\right)^\wedge_{1k}=0,\qquad k=1,2,3,4,
\label{f-2}\eeq
give the following polynomial $B(u)$ for the separation 
variables $u_1$ and $u_2$:
\beq 
B(u)=u^4+B_2u^2+B_0=(u^2-u_1^2)(u^2-u_2^2),
\label{f-3}\eeq
\beq
B_2=\frac{2b(J_3+c)(x_3J_--x_-J_3)+b^2(x_-^2+2x_3^2)}{J_-^2+2bx_-}
+\frac{2b^2x_3x_-(cJ_-+bx_3)}{J_-^2(J_-^2+2bx_-)}\,,
\label{f-4}\eeq
\beq
B_0=b^2\,\frac{\left((J_3+c)J_-+bx_3\right)^2}{J_-^2(J_-^2+2bx_-)}\,.
\label{f-5}\eeq
Note also the following useful formula for the variable $B_2$:
\beq
B_2=\frac{b^2x_-^2}{J_-^2+2bx_-}-\frac{2\sqrt{B_0}}{aJ_-\sqrt{J_-^2+2bx_-}}\,
\left(x_-J_-J_3-(J_-^2+bx_-)x_3\right),
\label{f-6}\eeq
which is a linear expression in terms of $J_3$ and $x_3$.

The corresponding rational function $A(u)$ can be chosen as follows:
\beq
A(u)=\frac{A_{-1}}{u}+\frac{A_{-3}}{u^3}\,,
\label{f-7}\eeq
\beq
A_{-1}=\mbox{\rm i}c+\frac{\mbox{\rm i}x_3(J_-^2+2bx_-)}{x_-J_-}\,, \qquad 
A_{-3}=\mbox{\rm i}b\,\frac{(J_3+c)J_-+bx_3}{x_-J_-}\,.
\label{f-8}\eeq

Now, it is easy to check the following Poisson brackets between
the two functions:
\beq
\matrix{\{A(u),A(v)\}=\{B(u),B(v)\}=0,\cr\cr
\{A(u),B(v)\}=\frac{2}{u^3(u^2-v^2)}\,\left(u^4B(v)-v^4B(u)\right),}\qquad \forall u,v\in\C.
\label{f-9}\eeq
Using these formulas one derives that the variables $u_1$, $u_2$, defined
by (\ref{f-3})--(\ref{f-6}), and their conjugated counterparts,
\beq
v_j=A(u_j),\qquad j=1,2,
\label{f-10}\eeq
are indeed canonical (Darboux) variables:
\begin{equation}
\{u_j,u_k\}=\{v_j,v_k\}=
\{u_j,v_k\}=0,\qquad j\neq k,\qquad \{v_j,u_j\}=1,\qquad j=1,2.
\label{ha}
\end{equation}
These variables by construction satisfy (for $\ell=0$) to (\ref{f-2}) and,
therefore, to (\ref{cucu-4}). 

Notice here that the definitions
of the separation variables do not depend on the value of $\ell$
and also that the brackets (\ref{ha}) are true {\it for any $\ell$}.

\section{Generating function}

Let us fix a special representation of the 
underlying e(3) algebra which will allow us to write down
the found canonical transformation explicitly through
the generating function. It is a kind of holomorphic
representation in terms of the Darboux variables
$q_j$, $p_j$, $j=1,2$:\footnote{cf. a holomorphic representation
of sl(2)}
\beq
J_-=q_1,\qquad x_-=q_2,
\label{f-11}\eeq
\beq
J_3=\mbox{\rm i}(q_1p_1+q_2p_2)+\ell,\qquad x_3=\mbox{\rm i}q_2p_1+1,
\label{f-12}\eeq
\beq
J_+=q_1p_1^2+2q_2p_1p_2-2\mbox{\rm i}\ell p_1-2\mbox{\rm i}p_2,\qquad
x_+=q_2p_1^2-2\mbox{\rm i}p_1.
\label{f-13}\eeq

The meaning of this realization is in the fact that the variables $J_-$ and 
$x_-$ do not depend on the momenta ${\bf p}$ and the variables $J_3$
and $x_3$ are linear in momenta. This together with linearity
of the variables $u_1u_2$ and $u_1^2+u_2^2$ in terms of
$J_3$ and $x_3$ make it possible to integrate the equations explicitly
in terms of elementary functions.

Finally, the canonical transformation defined in the previous Section
is given by the following generating function $F({\bf u}|{\bf q})$:
\be
F({\bf u}|{\bf q})=\frac{\mbox{\rm i}q_1}{q_2}+\frac{\mbox{\rm i}c}{2}\,
\log(q_1^2+2bq_2)+\mbox{\rm i}\ell\log(q_2)+\frac{\mbox{\rm i}b^2q_2}
{2u_1u_2\sqrt{q_1^2+2bq_2}}
\nonumber\ee
\beq
+\;\frac{\mbox{\rm i}}{2}\,\sqrt{q_1^2+2bq_2}\left( 2\,\frac{u_1u_2}{b} 
-\frac{u_1^2+u_2^2}{q_2u_1u_2}\right).
\label{f-15}\eeq
Therefore, the equations of the change of variables $({\bf q},{\bf p}) \leftrightarrow
({\bf u},{\bf v})$ are written in the form
\beq
p_j=\frac{\partial F({\bf u}|{\bf q})}{\partial q_j}\,,\qquad
v_j=-\frac{\partial F({\bf u}|{\bf q})}{\partial u_j}\,,\qquad j=1,2.
\label{f-16}\eeq

\section{Inverse problem}
The inverse problem, i.e. finding expressions for the initial
e(3) variables in terms of separation variables, also has 
an explicit solution which is given below:
\beq
q_1^2=-b^2\,\frac{u_1^2(u_2^4v_2^2-b^2)-u_2^2(u_1^4v_1^2-b^2)}
{4u_1^6u_2^6(u_1v_1-u_2v_2)^2}\,
\left( u_1^2(u_2^4v_2^2-b^2)-u_2^2(u_1^4v_1^2-b^2)+4u_1^2u_2^2(u_1^2-u_2^2)\right),
\label{f-17}\eeq
\beq
J_-=q_1,\qquad
x_-=\frac{b(u_1^2-u_2^2)\left(u_1^2(u_2^4v_2^2-b^2)-u_2^2(u_1^4v_1^2-b^2)\right)}
{2u_1^4u_2^4(u_1v_1-u_2v_2)^2}\,,
\label{f-19}\eeq
\be
J_3=\frac{1}{2u_1^2u_2^2(u_1v_1-u_2v_2)
\left(u_1^2(u_2^4v_2^2-b^2)-u_2^2(u_1^4v_1^2-b^2)\right)}\;
\times \qquad \qquad\qquad \qquad
\nonumber\ee
\be
\; \times\;\left(  \;\mbox{\rm i}b^4(u_1^2-u_2^2)^2
+2cu_1^4u_2^4(u_1v_1-u_2v_2)\left(u_1^2(v_1^2-2)-u_2^2(v_2^2-2)\right) \right. 
\qquad \qquad
\qquad \qquad
\nonumber\ee
\be
\left. +\;2cb^2u_1^2u_2^2(u_1v_1-u_2v_2)(u_1^2-u_2^2) 
+ 2\mbox{\rm i}
b^2u_1^2u_2^2(u_1^2v_1^2-u_2^2v_2^2)(u_1^2-u_2^2) 
 \right.\qquad \qquad
\nonumber\ee
\beq
\left. +\;\mbox{\rm i}u_1^4u_2^4(u_1v_1-u_2v_2)
\left(v_1u_1^3(v_1^2-4)-v_2u_2^3(v_2^2-4)+u_1u_2v_1v_2(u_1v_1-u_2v_2)\right) 
 \;\right),
\label{f-23}\eeq
\beq
x_3=\frac{2\mbox{\rm i}q_1u_1^2u_2^2\left(u_1^2(u_1v_1-\mbox{\rm i}c)-u_2^2(u_2v_2-
\mbox{\rm i}c\right)}{b\left( u_1^2(u_2^4v_2^2-b^2)-u_2^2(u_1^4v_1^2-b^2)\right)}\,,
\label{f-24}\eeq
\be
J_+=-\frac{4q_1u_1^4u_2^4(u_1v_1-u_2v_2)}
{b^2\left(u_1^2(u_2^4v_2^2-b^2)-u_2^2(u_1^4v_1^2-b^2)\right)^2}\;
\times \qquad \qquad\qquad \qquad\qquad \qquad\qquad 
\nonumber\ee
\be
\; \times\;\left(  \;u_1^3u_2^3v_1v_2(u_1v_1-u_2v_2)
+b^2(u_1^3v_1-u_2^2v_2)
-\mbox{\rm i} c b^2(u_1^2-u_2^2)\right.
\nonumber\ee
\beq
\left. -\;\mbox{\rm i}cu_1^2u_2^2
(u_1^2v_1^2-u_2^2v_2^2)-c^2u_1^2u_2^2(u_1v_1-u_2v_2) 
\;\right), 
\label{f-27}\eeq
\be
x_+=\frac{1}
{b\left(u_1^2(u_2^4v_2^2-b^2)-u_2^2(u_1^4v_1^2-b^2)\right)^2}\;
\times \qquad \qquad\qquad \quad
\qquad\qquad 
\qquad\qquad 
\nonumber\ee
\be
\; \times\;\left(\;-4\mbox{\rm i}cb^2u_1^2u_2^2(u_1^2-u_2^2)
(u_1^3v_1-u_2^3v_2)
+2b^2u_1^2u_2^2(u_1^2-u_2^2)(u_1^4v_1^2-u_2^4v_2^2)
\right.\quad 
\nonumber\ee
\be
\left. -2b^2c^2u_1^2u_2^2(u_1^2-u_2^2)^2
-2c^2u_1^4u_2^4(u_1^2-u_2^2)\left(u_1^2(v_1^2-4)
-u_2^2(v_2^2-4)\right)\right.
\nonumber\ee
\be
\left.
-\;4\mbox{\rm i}cu_1^4u_2^4(u_1^3v_1-u_2^3v_2)\left(
u_1^2(v_1^2-4)-u_2^2(v_2^2-4)\right)\qquad 
\qquad \qquad\qquad
\right. 
\nonumber\ee
\beq
\left.
+\;2u_1^4u_2^4\left(
(u_1^2v_1^2-u_2^2v_2^2)(u_1^4v_1^2-u_2^4v_2^2)
-4(u_1^3v_1-u_2^3v_2)^2\right)
\;\right).\qquad \qquad 
\label{f-32}\eeq

\section{Concluding remarks}
A separation of variables for the Kowalevski gyrostat, with $\ell=0$,
is found for the first time. It appeared to separate the Goryachev-Chaplygin 
gyrostat as well, thereby giving another separation for this problem.

If we put $c=0$, i.e. switch off the gyrostatic term, we can compare 
our results with the original Kowalevski's
and Goryachev-Chaplygin's separations for the respective tops
(in the $\ell=0$ case).
It is easy to see that the new separation is as complicated 
as the (simple) Goryachev-Chaplygin separation and it is {\it much
simpler} than the Kowalevski separation. 

Hence, the new separation stands a good chance to be quantized.

%
%%%%%%%%%%%%%%%%%%%%%%%%%%%%%%%%%%%%%%%%%
%
\section*{Acknowledgements}
%I wish to thank Ritchie Blackmore, Noddy Holder and Ozzy Osbourne 
%for the inspiration.
The support of the EPSRC is gratefully acknowledged.
%\pagebreak

%
%%%%%%%%%%%%%%%%%%%%%%%%%%%%%%%%%%%%%%%%%
%
\def\cprime{$'$}

\end{document}